\documentclass[twocolumn,showpacs,preprintnumbers,amsmath,amssymb,superscriptaddress]{revtex4}
\usepackage{amssymb}

\usepackage{amsmath,amssymb,graphicx,bm}

\begin{document}


\title{Modified clock inequalities and modified black hole lifetime}

\author{Rong-Jia Yang(ÑîÈÙ¼Ñ)}
\email{yangrj08@gmail.com}
 \affiliation{College of Physics and Technology,
Hebei University, Baoding 071002, China}
\author{Shuang Nan Zhang(ÕÅË«ÄÏ)}
 \email{zhangsn@mail.tsinghua.edu.cn}
 \affiliation{Department of Physics and Center for Astrophysics,
Tsinghua University, Beijing 100084, China} \affiliation{Key
Laboratory of Particle Astrophysics, Institute of High Energy
Physics, Chinese Academy of Sciences, P.O. Box 918-3, Beijing
100049, China} \affiliation{Physics Department, University of
Alabama in Huntsville, Huntsville, AL 35899, USA}

\date{\today}

\begin{abstract}
Based on a generalized uncertainty principle, Salecker-Wigner
inequalities are modified. When applied to black holes, they give a
modified black hole lifetime: $T_{\rm MB}\sim\frac{M^3}{m^3_{\rm
p}}(1-m^2_{\rm p}/M^2)t_{\rm p}$, and the number of bits required to
specify the information content of the black hole as the event
horizon area in Planck units: $N\sim\frac{M^2}{m^2_{\rm
p}}(1-m^2_{\rm p}/M^2)$.

\end{abstract}

\maketitle
\section{Introduction}
The conventional derivation of the Hawking lifetime uses the
Heisenberg's uncertainty principle on the event horizon scale
$R_{\rm g}$ to determine a temperature for the black hole which,
under the assumption that the black hole is a black body, then
allows one to use the Stefan-Boltzmann law to calculate the lifetime
of the black hole for complete evaporation (see, e.g.,
\cite{Townsend,Adler}).

By applying Salecker-Wigner's clock inequalities to black holes,
Barrow obtained the same result \cite{Barrow}. The heuristic way is
as follows: According to Heisenberg's uncertainty principle: $\Delta
p\sim\hbar/\Delta x$, if a clock of mass $M$ has quantum position
uncertainty $\Delta x$, then its momentum uncertainty is
$\hbar\Delta x^{-1}$. The clock to be considered should have an
accuracy $\tau$ (the minimum time interval that the clock is capable
of resolving) and be able to measure time intervals up to a maximum
$T$. After a time $t$, the uncertainty in position of the clock will
grow to $\Delta x'=\Delta x+\hbar tM^{-1}\Delta x^{-1}$. If the
effects on mass are neglected, then this will be a minimum when
$\Delta x=\sqrt{\hbar t/M}$. Hence, to keep the clock accurate over
the total running time $T$, its linear spread $\lambda$ must be
limited:
\begin{eqnarray}
\label{1}\lambda\geq2\sqrt{\hbar T/M},
\end{eqnarray}
the same order of magnitude of the position uncertainty, meaning
that the size of the clock must be larger than the uncertainty in
its position. This is Salecker-Wigner's first clock inequality
\cite{Wigner}. To give time to within an accuracy $\tau$, the
quantum position uncertainty must not be larger than the minimum
wavelength of the quanta striking it (in order to read the time);
that is, $\Delta x'\leq c\tau$. The use of a signal with nonzero
rest mass would give a more rigorous limit. This condition gives a
bound on the minimum mass of the clock:
\begin{eqnarray}
\label{2}M\geq\frac{4\hbar}{c^2\tau}\left(\frac{T}{\tau} \right).
\end{eqnarray}
This is Salecker-Wigner's second clock inequality \cite{Wigner}.
This inequality is more restrictive than that imposed by
Heisenberg's energy-time uncertainty principle because it requires
that a clock still show proper time after being read: the quantum
uncertainty in its position must not introduce significant
inaccuracies in its measurement of time over the total running time.
To derive Salecker-Wigner's clock inequalities (\ref{1}) and
(\ref{2}), it assumes unsqueezed, unentangled, and Gaussian wave
packets without any detailed phase information; they are valid only
for single analog clocks (black holes can be seen as analog clocks
\cite{Ng01}), not for digital quantum clocks.

Barrow applied Salecker-Wigner's size limit (\ref{1}) to a black
hole, assuming that the minimum clock size is the Schwarzschild
radius $R_{\rm g}=2GM/c^2$ and found the maximum running time of the
black hole is \cite{Barrow}:
\begin{eqnarray}
\label{4}T\sim\frac{G^2M^3}{\hbar c^4}=\frac{M^3}{m^3_{\rm p}}t_{\rm
p},
\end{eqnarray}
where $t_{\rm p}\equiv\sqrt{G\hbar/c^5}$ and $m_{\rm
p}\equiv\sqrt{c\hbar/G}$ are the Planck time and mass. The maximum
running time of a black hole is the Hawking lifetime \cite{Hawking}.
If we had not known of the existence of black hole evaporation, Eq.
(\ref{4}) would have implied that there is a maximum lifetime for a
black hole state. Compared with the conventional method, the
application of the Salecker-Wigner inequality (\ref{1}) to the event
horizon scale predicts the Hawking lifetime (\ref{4}) without the
assumption that the black hole is a black body radiator.

But, one may suggest, when considering black holes, the effect of
gravity may be taken into account. In this work, we obtain modified
clock inequalities based on a generalized uncertainty principle that
takes into account some properties of black holes, and find a
modified black hole lifetime which may throw light on quantum
gravity at the Planck scale.

\section{Modified clock inequalities}
Salecker-Wigner's clock inequalities are based on the Heisenberg's
position-momentum uncertainty principle: $p\sim\hbar/\Delta x$. But,
if we combine quantum theory and some basic concepts of gravity,
Heisenberg's position-momentum uncertainty principle may be modified
\cite{Veneziano, Gross, Amati, Konishi, Guida, Kato, Maggiore,
Witten, Garay, Bambi, Park, Kim08, Machluf,
Nath,Capozziello,Camacho,Bina,Castro}, and so do Salecker-Wigner's
clock inequalities. Using Heisenberg's uncertainty principle and
some properties of black holes, Scardigli had shown how a
generalized uncertainty principle (GUP) can be derived from a
measure gedanken experiment \cite{Scardigli}:
\begin{eqnarray}
\label{11}\Delta x\geq\frac{\hbar}{\Delta p}+l^2_{\rm p}\frac{\Delta
p}{\hbar},
\end{eqnarray}
where $l^2_{\rm p}=\sqrt{G\hbar/c^3}$ is the Planck distance. As
Scardigli argued, this GUP is independent from particular versions
of quantum gravity. This GUP also arises from quantum fluctuation of
the background space-time metric \cite{Adler99}. Note, however, this
GUP is firstly derived in Ref. \cite{Maggiore}. The GUP (\ref{11})
can be written in a general form $\Delta x\geq \hbar(1/\Delta
p+\beta \Delta p)$, where $\beta$ is a constant \cite{Chang}.
Introduction of the GUP has drawn considerable attention and many
authors considered various problems in the framework of GUP, such as
Refs.
\cite{Medved,Ashoorioon,Vakili,Bolen,Nozari,Kim,Ahluwalia,Nasseri,Maziashvili,
Myung,Akhoury,Doplicher,Kalyana,Hossenfelder,Scardigli03,Kim09,Larranaga,
Liu,Sakhawat,Valerio,Bang,Shibusa,Matsuo,Cust,Setare,Das,Ko,Sarris,li,Nouicer,Arraut,Galan,Alexander}.
Note, however, it should be kept in mind that this GUP is derived
based upon only heuristic arguments, and is thus far from proven.

Basing on the GUP (\ref{11}), Adler \emph{et al.} obtained a
modified black hole lifetime with the conventional method
\cite{Adler}.
\begin{eqnarray}
\label{ad} T_{\rm
ACS}=&&\frac{1}{16}\Big\{\frac{8}{3}\Big(\frac{M}{m_{\rm
p}}\Big)^3-8\frac{M}{m_{\rm p}}-\frac{m_{\rm
p}}{M}+\frac{8}{3}\Big[\Big(\frac{M}{m_{\rm p}}\Big)^2-1\Big]^{3/2}
\nonumber\\&&-4\sqrt{\Big(\frac{M}{m_{\rm
p}}\Big)^2-1}+4\arccos\Big(\frac{m_{\rm
p}}{M}\Big)+\frac{19}{3}\Big\}t_{\rm ch}
\end{eqnarray}
where $t_{\rm ch}=16^2\times60\pi t_{\rm p}$. To derive this black
hole lifetime, Adler \emph{et al.} also assume that the black hole
is a black body radiator and the dispersion relation $E=pc$ holds.
But if the uncertainty principle is modified, the dispersion
relation may also be modified (see, e.g., \cite{Nozari06}).

Because the space-time fluctuation will be significant when the
measured length scale approaches to the Planck distance, it is
reasonable to expect that the linear spread of a clock must not be
less than the Planck distance. In fact, the GUP (\ref{11}) implies a
minimum length: $2l_{\rm p}$, which can be considered as a limit on
the linear spread of a clock. This limit can be improved, as we see
below. From Eq. (\ref{11}), if a clock of mass $M$ has quantum
position uncertainty $\Delta x$, then its momentum uncertainty will
be $\Delta p\sim\frac{\Delta x\hbar}{2l^2_{\rm
p}}\left[1-\sqrt{1-4l^2_{\rm p}/\Delta x^2}\right]$ \cite{Adler}.
Following the steps to derive the Salecker-Wigner's clock
inequalities, Eq. (\ref{1}) is modified as (see Appendix)
\begin{eqnarray}
\label{13}\lambda\geq 2l_{\rm p}\sqrt{1+\frac{\hbar T}{Ml^2_{\rm
p}}},
\end{eqnarray}
stronger than limit (\ref{1}) and come back to limit (\ref{1}) for
$\hbar T\gg Ml^2_{\rm p}$. Here we also require that the position
uncertainty created by the measurement of time must not be larger
than the minimum wavelength of the quanta used to read the clock.
Then Salecker-Wigner's second clock inequality (\ref{2}) is modified
as:
\begin{eqnarray}
\label{20}M\geq \frac{4\hbar T}{c^2\tau^2}\frac{1}{1-4t^2_{\rm
p}/\tau^2},
\end{eqnarray}
This inequality links the mass, total running time, accuracy of the
clock, and the Planck time together, and may links together our
concepts of gravity and quantum uncertainty. Obviously, it firstly
gives a limit on the accuracy of the clock $\tau>2t_{\rm p}$. Like
Salecker-Wigner inequalities (\ref{1}) and (\ref{2}), Eqs.
(\ref{13}) and (\ref{20}) are valid for single analog clocks, not
for digital quantum clocks.

\section{Modified black hole lifetime}
Now applying modified clock inequality (\ref{13}) to black holes and
assuming that the minimum clock size is the Schwarzschild radius
$R_{\rm g}=2GM/c^2$, one may find the maximum running time of the
black hole is modified as:
\begin{eqnarray}
\label{22}T_{\rm MB}\sim\frac{MR^2_{\rm g}}{4\hbar}(1-4l^2_{\rm
p}/R^2_{\rm g})=\frac{M^3}{m^3_{\rm p}}(1-m^2_{\rm p}/M^2)t_{\rm p},
\end{eqnarray}
which has a term $Mt_{\rm p}/m_{\rm p}$ different from the Hawking
lifetime (\ref{4}) and holds for $M\geq m_{\rm p}$. This difference
may throw light on quantum gravity in some sense at Planck scale.
Using the GUP (\ref{11}), Adler \emph{et al.} found that the thermal
radiation of the black hole will stop at the Planck distance, and
the black hole becomes an inert remnant, possessing only
gravitational interaction \cite{Adler}, consistent the results
obtained in modified clock inequalities background. Aside from about
a factor of $16^2\times60\pi$, the first two terms of the
Adler-Chen-Santiago lifetime $T_{\rm ACS}$ is consistent with the
modified black hole lifetime $T_{\rm MB}$. The comparison among the
Hawking lifetime $T_{\rm H}$, the modified black hole lifetime
$T_{\rm MB}$, and Adler-Chen-Santiago lifetime $T_{\rm ACS}$ are
shown in Fig. $1$.
\begin{figure}
\includegraphics[width=9cm]{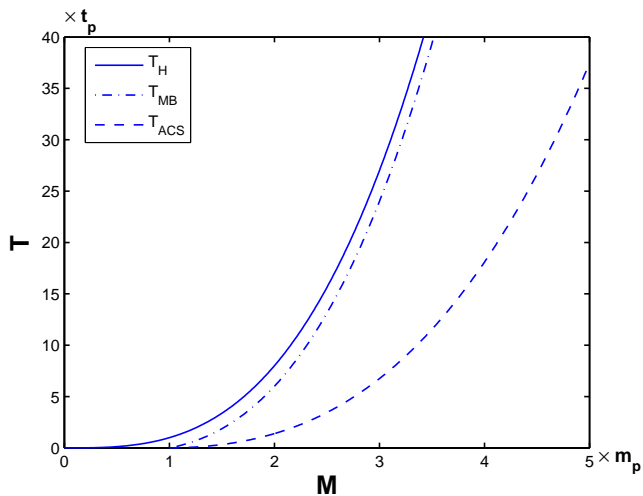}
\caption{Comparison among the Hawking lifetime $T_{\rm H}$, modified
clock inequality lifetime $T_{\rm MB}$, and Adler-Chen-Santiago
lifetime $T_{\rm ACS}$, aside from a numerical factor
$16^2\times60\pi$. \label{Fig1}}
\end{figure}

The minimum interval that the black hole can be used to measure is
given by the light travel time across the black hole \cite{Barrow,
Ng01}: $\label{06}\tau\sim2GM/c^3=R_{\rm g}/c$. Thus we are led to
view the black hole as an information-processing system in which the
number of computational steps is
\begin{eqnarray}
\label{22}N\equiv\frac{T_{\rm MB}}{\tau}\sim\frac{M^2}{m^2_{\rm
p}}(1-m^2_{\rm p}/M^2).
\end{eqnarray}
As expected from the identification of a black hole entropy
\cite{Bekenstein} or holographic principle \cite{Hooft,Susskind},
this gives the number of bits required to specify the information
content of the black hole as the event horizon area in Planck units.

\section{Summary}
To summarize, based on a generalized uncertainty principle, we
obtain modified clock inequalities, which give bounds on the size
and the accuracy of the analog clock that must be lager than $2$
times the Planck distance $l_{\rm p}$ and time $t_{\rm p}$
respectively. As an application, we discussed the case of black
holes, and obtained a modified black hole lifetime $T_{\rm
MB}\sim\frac{M^3}{m^3_{\rm p}}t_{\rm p}(1-m^2_{\rm p}/M^2)$, which
is different from Hawking lifetime and give a limit on the mass of
black holes naturally. Viewing a black hole as an
information-processing system, we also find the number of bits
required to specify the information content of the black hole as the
event horizon area in Planck units $N\sim\frac{M^2}{m^2_{\rm
p}}(1-m^2_{\rm p}/M^2)$. These results reinforce the central
importance of black holes as the simplest and most fundamental
constructs of space-time, linking together our concept of gravity,
information, and quantum uncertainty. Note, however, applying clock
inequalities to obtain the lifetimes of other type black holes is
still an open interesting problem, work is in progress in this
direction.

\begin{acknowledgments}
China, Directional Research Project of the Chinese Academy of
Sciences under Project No. KJCX2-YW-T03 and by the National Natural
Science Foundation of China under project Nos. 10521001, 10733010,
10725313.
\end{acknowledgments}

\section*{Appendix}
According to the generalized uncertainty principle
\begin{eqnarray}
\label{a0}\Delta x\geq\frac{\hbar}{\Delta p}+l^2_{\rm p}\frac{\Delta
p}{\hbar},
\end{eqnarray}
where $l^2_{\rm p}=\sqrt{G\hbar/c^3}$ is the Planck distance, if a
clock with mass $M$ has quantum position uncertainty $\Delta x$,
then its momentum uncertainty will be
\begin{eqnarray}
\Delta p\sim\frac{\Delta x\hbar}{2l^2_{\rm
p}}\left[1-\sqrt{1-4l^2_{\rm p}/\Delta x^2}\right].
\end{eqnarray}
After a time $t$ the uncertainty in position of the clock becomes
\begin{eqnarray}
\label{a1} \Delta x'=\Delta x+\frac{\Delta x\hbar t}{2Ml^2_{\rm
p}}\left[1-\sqrt{1-4l^2_{\rm p}/\Delta x^2}\right].
\end{eqnarray}
To obtain the minimal value of $\Delta x'$ in this case, using the
condition
\begin{eqnarray}
\label{a2} 0&=&\frac{d\Delta x'}{d\Delta x}\\\nonumber
&=&1+\frac{\hbar t\left[1-\sqrt{1-4l^2_{\rm p}/\Delta
x^2}\right]}{2Ml^2_{\rm
p}}-\frac{2t\hbar}{M\Delta x^2\sqrt{1-4l^2_{\rm p}/\Delta
x^2}},
\end{eqnarray}
we get $\Delta x=[2Ml^2_{\rm p}+t\hbar]/\sqrt{M(Ml^2_{\rm
p}+t\hbar)}$. By inserting this value into Eq. (\ref{a1}), we obtain
the minimal value of $\Delta x'$
\begin{eqnarray}
\Delta x'_{\rm min}=2l_{\rm p}\sqrt{1+\frac{t\hbar}{Ml^2_{\rm p}}}.
\end{eqnarray}
By taking $t$ as the total running time $T$ during which the clock
can remain accurate, and consider the condition that the linear
spread of clock $\lambda$ must not be less than the uncertainty in
position $\Delta x'$, that's $\lambda\geq\Delta x'\geq\Delta x'_{\rm
min}$, we obtain Eq. (\ref{13}).

\bibliography{apssamp}

\end{document}